# Comment to: "The interaction of relativistic spacecrafts with the interstellar medium"

Marko Karlušić

Ruđer Bošković Institute, Bijenička 54, 10000 Zagreb, Croatia

Marko.karlusic@irb.hr

**Abstract**

Recently, Hoang et al. (arXiv:1608.05284) reported analysis of the interaction of relativistic spacecrafts with interstellar medium (ISM, i.e. gas atoms and dust particles) relevant for the Breakthrough starshot initiative (https://breakthroughinitiatives.org/Initiative/3). The main conclusion is that dust pose much greater threat to the starship than gas atoms. However, analysis used to treat interaction of the spaceship with gas atoms is based on the incorrect use of the Szenes model. Only by proper treatment of the Szenes model can be found if the conclusion remains valid - or not. In the following, the main comments we have raised about the paper are listed. Present text is based on the v2 version of the above mentioned paper [0] that was accepted for publication in Astophysical Journal.

1.   **VELOCITY EFFECT.** To describe interaction of the spacecraft with interstellar dust, authors have employed Szenes model (analytical thermal spike model, ATSM). The way Hoang and co-authors are using it (considering only energy loss dE/dx and critical energy density of track formation $S_{et}$) is not correct because it does not take into account the so-called "velocity effect". In short, identical dE/dx can be achieved for two different ion velocities (one to the left of the Bragg peak is low velocity point, and the other to the right is high velocity data point, Fig.2 in [0]). For some reasons (still under debate, for example see introduction in [1]) track formation is much easier at low velocities, and therefore threshold for ion track formation is lower. Therefore, use of experimental data from low velocity ion irradiation of quartz (figure 4a from [0]) is incorrect because the threshold for ion track formation at 4 keV/nm underestimates threshold for high velocity gas impacts from the ISM (could be as high as 8 keV/nm). Also, data from the figure 4b (graphite [0]) are not all from the high velocity regime. Furthermore, velocity effect was introduced into the paper only due to decrease of the dE/dx beyond Bragg peak at very highest velocities, but there are other physical mechanisms actively contributing to ion track formation at lower energies ([1] and references therein).

2.   **QUARTZ.** ATSM parametrization of $a_0$=2.2 nm (quartz) is not correct either. This material is an insulator and Szenes found that all insulators show uniform behavior with $a_0$=4.5 nm. Fitting done in the paper [0] is not correct because "subthreshold damage" is often found due to different damage production mechanisms. Most likely point like damage produced below the ion track threshold accumulates for larger fluences that at some point makes phase transition into amorphous phase [2]. For this reason, Szenes is careful not to analyze very small ion tracks because RBS/c data (like the one in fig. 4a [0]) measure ion tracks indirectly, and it is not easy to distinguish small ion track close to the threshold from the "subthreshold damage build-up". While the quartz was used as an example to show how ATSM does not describe ion track formation well [3], recent study in our opinion validates ATSM [4]. Besides small amount of damage below 4 keV/nm, overall ion track evolution seems to be well described by $a_0 = 4.5$ nm. Also we note there is Szenes' analysis of Meftah's data [5].

3. **GRAPHITE.** Regarding graphite (or HOPG, highly oriented pyrolytic graphite), there are several problems. First, detailed studies are done only by STM and Raman spectroscopy. STM observes only the surface of the material, while Raman is difficult to interpret although there is a lot of ongoing work in this direction because of similarities between graphene and graphite. But in both cases (STM and Raman studies on graphene) it is shown that close to the threshold not every projectile produces an ion track. For example (STM data on HOPG, figure 4b [0]), threshold is reported at 7 keV/nm but 100% track forming efficiency is found only at 18 keV/nm. This should have consequence for the further analysis (surface and volume damage). Second problem is a general problem with the experimental data obtained by the surface probing techniques like STM and AFM. Due to the tip-size effects, it is difficult to measure the ion track size directly, so this type of data is usually not used for the quantitative analysis. Third problem, is that HOPG is very special, highly anisotropic material, and all models (not just ATSM) have to use non-standard parameters to describe ion tracks in it. Our estimate was that parameter $a_0 = 2-3$ nm should be used for HOPG [6]. Still, it remains an open question what is exactly threshold for ion track formation (where they appear or where the efficiency = 1), with all the consequences for the fit on the figure 4b [0]. We note that parametrization Hoang et al. [0] have used for hopg ($\alpha = 0.2$) resembles ATSM equations but it is actually not allowed within ATSM.

4. **ATSM.** The Szenes model for ion track formation [7] uses two model parameters, initial width of the thermal spike $a_0$ and fraction of deposited energy transferred into the thermal spike g. For insulators these two parameters are fixed, i.e. $a_0 = 4.5$ nm, and g = 0.4 or g = 0.17. For the low velocity ions, i.e. for specific kinetic energy below E/A = 2 MeV/amu, g = 0.4, while for high velocities, i.e. above E/A = 8 MeV/amu, g = 0.17. For intermediate energies, this fraction g varies depending on the material. Threshold for the ion track formation is given by equation (3) from ref. [7], and to be consistent one has to have this calculated $S_{et}$ equal to the $S_{et}$ obtained experimentally. Furthermore, ion track evolution is logarithmic-like close to the threshold and power-like further away, eqs. (2a, 2b) from ref. [7].

In order to describe ion tracks in the material on board of the spacecraft (v = 0.2c corresponds to 18 MeV/amu), one has to take into account all of this because ion track does not have uniform profile (cross section). As soon as the gas atom impacts the surface of the material, it starts to decelerate, and therefore its energy loss (stopping power dE/dx) changes: first it increases up to the maximum (Bragg peak) then it falls down again. But when the velocity effect is taken fully into account, one can see that ion tracks can be formed in the bulk, even when there are no tracks on the surface! Therefore, high velocity movement of the spacecraft does not protect it from ion track damage, as discussed in section 7 [0]. This is because thresholds for ion track formation are not the same at high and low velocities, due to the g parameter of the ATSM. This should have significant consequences to the volume factors that have been calculated [0], because full ion track profile along the whole ion trajectory should be taken into account.

5. **SPUTTERING.** As a thermally driven process, sputtering by gas atoms can also take place. Szenes treated SiO$_2$ sputtering in ref. [8] that is not related to sublimation of the material. For dust bombardment, Hoaeng et al. estimate melting and vaporization tempeartures [0]. For these nanoimpactors, nonlinear effects should be important, hence validity of eq. (11) from ref. [0] is questionable. Although there is still debate if ion tracks formed by cluster ions (like C$_{60}$) can be compared to the monoatomic tracks or not, for larger grains such approach (using stopping power) should not hold. Perhaps evaporative sputtering from the dust particles is not all bad news, in a sense that it removes damaged material. It also might be worthwhile checking if ion tracks produced before dust impact do sensitize material to these processes. Finally, there can be sputtering (and accompanying ion track formation) taking place from cosmic rays, unless their abundance is too low for significant effect [9].

6. **OTHER COMMENTS.**
   (a) Number of gas atoms to the α-Centauri is $2*10^{18}$ /cm$^2$. For 20 year trip, neglecting relativistic effects, this amounts to constant flux of $2*10^9$ impacts/cm$^2$/s. This is typical flux that can be achieved in high energy ion accelerator, and for 18 MeV protons this corresponds to cca. 5 mW heating for 1x1 cm$^2$ surface area. Generally it is considered that only tens of degrees heating of the target is achieved due to such energy deposition. For this reason we find it suspicious that temperature increases authors have calculated [0] are of the order of few hundred degrees, although the starship is perfect thermally insulated system. We note also that most of the track studies have been done at room temperature. Going to much lower temperatures (like in the space) should increase threshold for ion track formation. On the other hand, for excess temperatures, production of ion tracks can be facilitated (but tracks can also be erased, i.e. annealed).
   (b) There can be also charging of the starship due to gas impacts. It is known that impacts of heavy MeV ions can result in large secondary electron release (hundreds of electrons). For $10^{16}$ impacts (assuming 1:100 heavy ion/proton ratio), there can be $10^{18}$ electrons missing. This charging can have adverse effects to the ship electronics. Similarly, cosmic rays can have negative impact on ship electronics that has to be looked into.
   (c) Catastrophic impact (complete evaporation of ship due to large dust impact) was estimated to be extremely unlikely. It would be of great importance to estimate (from distribution of dust particle sizes) what is the critical size of the dust that has total impact probability of 1% or 10%? This way, one has to consider only impacts of dust particles with smaller size, assuming smaller risks related to larger particles can be mitigated in other ways (for example by sending multiple ships).
   (d) Ion channeling can be important issue in reducing damage to the starship. Assuming gas atoms to be stationary, and given high velocity of the ship, almost all impactors will be channeled if material (single crystal quartz or HOPG) is oriented precisely in the direction of the ship movement. When gas atom undergo channeling, their stopping power can be reduced 30-50%, and damage to the surface will be less severe.
   (e) Ion tracks will be largest in the material (not on the surface), and depending on the accumulated fluence (and ion type), these amorphous pockets will exert pressure on the material, resulting in its swelling. For higher fluences, material will become more and more damaged (amorphous material is more sensitive to ion track formation), and synergistic effects can take place, even leading to formation of bubbles or voids,

- (f) resulting in reduction of materials mechanic and thermal properties. Furthermore, even for protons and helium, nuclear energy loss at the end of the range can also cause a lot of damage (at these fluences) – for example "smart cut" is the technological way to produce thin films by high fluence ($10^{17}$ cm$^{-2}$) irradiation and subsequent delamination.
- (f) While insulators are susceptible to ion track formation, semiconductors and metals are less sensitive (or not at all). Metals might be too heavy to use, but semiconductors like Si and SiC could be attractive candidates. Not only are these materials very radiation hard (ion tracks are found only under very extreme conditions like cluster ion irradiation), they are also very good in terms of healing: irradiation by MeV heavy ions leads to erasing of existing damage [10]. It is still unclear exact mechanism of this recovery, weather it is recrystallization of the track during the cooling phase, or recombination of defects after thermal spike produced by nearby ion impact, but still could be interesting defensive mechanism. Also there are other strategies like use of amorphous materials (it's difficult to damage damaged material), that are investigated in the nuclear waste immobilization studies. Here we only want to stress that picking up the right material in the beginning (in terms of radiation hardness) could be very beneficial to the final starship design. Unfortunately, nuclear stopping damage is not avoidable by these strategies (except by utilizing ion channeling).
- (g) The ATSM model is only one of the models. While the thermal spike models describe ion track formation rather well, there are concerns about their validities. In a study like this, in our opinion one should not be concerned about these issues. For the risk assessment of the starship voyage, simply looking for the worst case scenario by using all of the available models should be sufficient.
- (h) The experiment would be interesting. Actually, given gas atom fluence and starship velocity, there is no accelerator facility that can deliver needed ion beams. Largest facilities that can provide 10 MeV/amu beams, typically have low ion fluxes ($10^9$-$10^{10}$ ions/cm$^2$/s), and those that can reach needed fluences can do so only at MeV energies (total kinetic energy). Medium-sized accelerator facilities could be even the best choice. Assuming 1:100 abundance of heavy elements compared to the protons, this would mean $10^{15}$-$10^{16}$ fluence that is in the range for heavy ion beams with energies below 100 MeV. Of course, these ion beams cannot reach v = 0.2c, but most of the damage is done at velocities below Bragg peak anyway. Furthermore, we have established that Bragg peak and related changes in ion track formation due to velocity effect are substantially below 2 MeV/amu for lighter ions like oxygen [11].

To conclude, our greatest concern is misuse of the experimental data that serve as input parameters for the analytical thermal spike model - ATSM. For the ion tracks in quartz, these low velocity data points have been erroneously used for high velocity study, without taking into account velocity effect. For the graphite, experimental data are not only from high velocity regime (there is also an open question about velocity effect for the surface tracks [6]), but more of concern is missing ion track yield in calculation of surface coverage (eq. (5) in ref [0]). Finally, there is a problem with the volume filling factor because ion tracks do not have constant profile along ion trajectories. Until these issues are clarified, conclusion that the damage to the starship from the interstellar gas is of less concern [0], might be underestimated.

# References

[0] T. Hoang, A. Lazarian, B. Burkhart, A. Loeb, The interaction of relativistic spacecrafts with the interstellar medium ([v2] Tue, 27 Dec 2016), https://arxiv.org/abs/1608.05284v2

[1] M. Karlusic et al., https://arxiv.org/abs/1606.03870

[2] F. Agulo-Lopez et al., Prog. Mat. Sci. 76 (2016) 1

[3] M. Toulemonde et al., NIMB 277 (2012) 28

[4] O. Pena-Rodriguz et al., J. Nucl. Mater 430 (2012) 125

[5] G. Szenes, NIMB 122 (1997) 530

[6] M. Karlusic et al., NIMB 280 (2012) 103

[7] G. Szenes, Phys. Rev. B 51 (1995) 8026

[8] G. Szenes, NIMB 233 (2005) 70

[9] G. Szenes et al., ApJ 708 (2010) 288

[10] A. Benyagoub et al., Appl. Phys. Lett. 89 (2006) 241914

[11] I. Bogdanovic-Radovic et al., Phys. Rev. B 86 (2012) 165316